\documentclass[aps,pra,amsmath,amssymb,amsfonts,superscriptaddress,floatfix,nofootinbib]{revtex4} 
\pdfoutput=1
\usepackage{amsmath,amsthm,amsfonts,amssymb}
\usepackage{graphicx}

\usepackage{color}
\usepackage{hyperref}

\usepackage{algorithm}
\usepackage[capitalize,nameinlink]{cleveref}

\newcommand{\be}{\begin{equation}}
\newcommand{\ee}{\end{equation}}

\newcommand{\pZB}{\partial_Z^B}
\newcommand{\ZZ}{\mathbb{Z}_2}
\newcommand{\ZZZZ}{\mathbb{Z}_4}
\newcommand{\cC}{{\cal C}}
\newcommand{\tcC}{\tilde{\cal C}}

\begin{document}

\title{A Sparse $\ZZ$ Chain Complex Without a Sparse Lift}
\author{Matthew B.~Hastings}
\begin{abstract}
We construct a sparse $\ZZ$ chain complex (with three different degrees, so that it corresponds to a quantum code) which does not admit a sparse lift to the integers, answering a question in Ref.~\cite{freedman2021building}.
\end{abstract}
\maketitle

A CSS (Calderbank-Shor-Steane) stabilizer quantum code corresponds to a chain complex $\cC$ with $\ZZ$ coefficients, and with vector spaces in three different degrees, where basis elements of these three vector spaces correspond to $Z$-stabilizers, qubits, and $X$-stabilizers respectively.  A code is said to be LDPC (``low-density parity check") whenever the boundary operators are sparse matrices, meaning that the rows and columns of the boundary operator have only $O(1)$ nonzero entries.  Here we are implicitly assuming the existence of a family of such codes where the number of qubits (corresponding to the dimension of the middle vector space) increases, but the number of nonzero entries remains bounded by some $O(1)$ constant.

One way to construct such quantum codes is by picking some cellulation of a manifold and picking some integer $d$ and letting $Z$-stabilizers, qubits, and $X$-stabilizers correspond to $(d+1)$-cells, $d$-cells, and $(d-1)$-cells respectively, using the boundary operator of the cellulation to define the chain complex.
If the cellulation has bounded local geometry, so that each cell is attached to a bounded number of other cells, then the resulting code is LDPC.

In Ref.~\cite{freedman2021building}, it was shown how to ``reverse engineer" a manifold from a code, taking as input an LDPC quantum code and outputting a cellulation of an $11$-manifold so that the manifold has bounded local geometry and so that for $d=4$ we can reconstruct the quantum code from that cellulation.  This reverse engineering procedure required, however, that the chain complex $\cC$ admit a \emph{sparse lift} $\tcC$.
A chain complex  $\tcC$ is said to be a lift of $\cC$ if $\tcC$ is a chain complex over the integers, whose boundary operator $\tilde \partial$ is equal, mod $2$, to the boundary operator $\partial$ of $\cC$.
A lift $\tcC$ is said to be a sparse lift if, for each row and column of $\tilde \partial$, the sum of absolute values of entries is $O(1)$.
In general, given two matrices, with one matrix $M_G$ having coefficients in some group $G$ and with another matrix $M_H$ having coefficients in some group $H$, with some homomorphism from $G$ to $H$, then $M_G$ is said to be a lift of $M_H$ if $M_H$ is the image of $M_G$ under this homomorphism.

It was shown\cite{freedman2021building} that every chain complex $\cC$ over $\ZZ$ admits some lift.  However, the question was left open as to whether every sparse chain complex admits a sparse lift.  Here we answer this question negatively.  Indeed, our chain complex has three degrees so it corresponds to some LDPC quantum code.

In \cref{Z4l} we introduce the general question of whether a $\ZZ$ chain complex admits a sparse lift to $\ZZZZ$.  Here, a lift to $\ZZZZ$ (or to any other finite group) is said to be sparse if each row and column of the boundary operator has $O(1)$ nonzero entries.  Indeed, if we can show that a chain complex admits no sparse lift to $\ZZZZ$, then it admits no sparse lift to the integers, as if it had a sparse lift to the integers then we could take the boundary operator modulo $4$ to define a sparse lift to $\ZZZZ$.
The reason for considering lifts to $\ZZZZ$ is that we can reduce this question to finding certain sparse solutions to equations over $\ZZ$.

In \cref{ex}, we introduce our example and in \cref{proof}, we show this example has no sparse lift.
Finally, \cref{locallift} we
address a question about lifting chain complexes which have certain locality properties with respect to a metric.  This question may be of some general interest
and also helps explain some features of our example.

\section{Lifting to $\ZZZZ$}
\label{Z4l}
We consider some chain complex with three degrees:
$$\cC=\ZZ^{n_Z} \stackrel{\partial_Z}{\rightarrow} \ZZ^{n_Q} \stackrel{\partial_Q}{\rightarrow} \ZZ^{n_X},$$
where $n_Z,n_Q,n_X$ are the dimensions of the three vector spaces.  This corresponds to the number of $Z$-stabilizers, qubits, and $X$-stabilizers in a quantum code.  Here we are using ``stabilizer" are shorthand for ``stabilizer generator".

We ask whether this chain complex admits some sparse lift to $\ZZZZ$.  Such a sparse lift means finding sparse matrices $\tilde \partial_Z$ and $\tilde \partial_Q$ with $\ZZZZ$ entries, which agree mod $2$ with $\partial_Z$ and $\partial_Q$, respectively, such that
\be
\label{Z4cc}
\tilde \partial_Q \tilde \partial_Z=0.
\ee

Following the result that every $\ZZ$ chain complex admits some lift to the integers, every chain complex admits some lift to $\ZZZZ$.  However, it is not obvious that it admits a sparse lift.  One possible lift is the ``naive lift", where we lift $0$ to $0$ and $1$ to $1$.  Such a lift gives sparse $\tilde \partial_Z$ and $\tilde \partial_Q$, however they may not be valid boundary operators as it is possible that the resulting matrices do not obey \cref{Z4cc}.

In this section, assume that we have some given lift of $\partial_Z$ to some matrix $L_Z$ and some lift of $\partial_Q$ to some matrix $L_Q$, meaning that these matrices agree mod $2$ with $\partial_Z$ and $\partial_Q$ respectively.
These $L_Z,L_Q$ need not be sparse.  Further, they need not obey $L_Q L_Z=0$; that is, they define some lift of the matrices $\partial_Z,\partial_Q$ but they do not necessarily define a lift of the chain complex.

Given these $L_Z,L_Q$, \emph{any} other matrices $L'_Z,L'_Q$ which define some other lift of $\partial_Z,\partial_Q$ must have
$L'_Z=L_Z \mod 2$ and $L'_Q=L_Q \mod 2$, so that
$$L'_Z=L_Z+2\eta_Z,$$
and
$$L'_Q=L_Q+2\eta_Q,$$
for some $\eta_Z,\eta_Q$.
Let $$L_Q L_Z=E_4,$$
for some matrix $E_4$.  Since $\partial_Q \partial_Z=0$, we must have $E_4=0 \mod 2$.

We have
\be
\label{defineeq}
L'_Q L'_Z=E_4+2L_Q \eta_Z + 2 \eta_Q L_Z.
\ee
This equation is mod $4$, so we still obtain equality if we replace $L_Q,L_Z,\eta_Z,\eta_Q$ by their values mod $2$.  That is, if we shift any entry of any of these matrices by $2$, the right hand side shifts by a multiple of $4$.
Let $\delta_Z,\delta_Q$ equal $\eta_Z,\eta_Q$ mod $2$, respectively.
Note that while \cref{defineeq} is an equation over $\ZZZZ$, every term in it is $0$ mod $2$.  So, we may obtain an equality over $\ZZ$ by dividing every term in \cref{defineeq} by $2$ and taking the equation mod $2$.
Let $E=E_4/2$, with entries of $E$ being in $\ZZ$.  Similarly regard $\delta_Z,\delta_Q$ as matrices with entries in $\ZZ$.
So,
$$
L'_Q L'_Z = 2(E +\partial_Q \delta_Z + \delta_Q \partial_Z).
$$
So, to have $L'_Q L'_Z=0$ we need
\be
\label{coneq}
E +\partial_Q \delta_Z + \delta_Q \partial_Z=0.
\ee
Here, \cref{coneq} is an equation over $\ZZ$.  The operators $\partial_Q,\partial_Z$ are simply the boundary operators of the original chain complex $\cC$.
The quantity $E$ is something that may be directly computed from the lift $L_Z,L_Q$.

Further, if $L_Z,L_Q$ is sparse, in order for a sparse $L'_Z,L'_Q$ to exist obeying $L'_Q L'_Z=0$, then $\delta_Z,\delta_Q$ must be sparse.
Thus, given any sparse choice of $L_Z,L_Q$, the question of whether the chain complex has a sparse lift to $\ZZZZ$ is equivalent to:
can we find a sparse choice of $\delta_Z,\delta_Q$ which solves  \cref{coneq} over $\ZZ$?

\section{Example}
\subsection{The Construction}
\label{ex}
Our example is actually a family of examples, controlled by a parameter $N$.  A quantity is $O(1)$ if it is bounded by a parameter independent of $N$.  We will, for brevity, say a quantity is ``large" if it is $\omega(1)$, i.e., it grows unboundedly with $N$.

The starting point of our example is a cellulation of the $4$-manifold $RP^3 \times [0,1]$.  We choose the cellulation to have bounded local geometry.  We subdivide the interval $[0,1]$ into some large number $N$ of subintervals, $[0,1/N],[1/N,2/N],\ldots$.  At each point $m/N$ in the interval for integer $m$, we have a cellulation of $RP^3$, and then there are additional cells connecting the cellulation of $RP^3$ at $m/N$ to that at $(m+1)/N$.  We choose the cellulation of $RP^3$ at $m=0$ to have size $O(1)$.  On the other hand, we choose the cellulations of $RP^3$ for $m>0$ to grow in size with increasing $m$.  Specifically, we choose them so that the shortest nontrivial $1$-cycle and shortest nontrivial $2$-cocycle each have a number of cells which grows as some unbounded function of $m$, and so that the total number of cells in at given $m$ grows as some unbounded function of $m$.  Thus, the shortest nontrivial $1$-cycle and shortest nontrivial $2$-cocycle of the cellulation of $RP^3$ at $m=N$ are both large.

Given this cellulation of $RP^3 \times [0,1]$, we define a quantum code that we call $B$.  This quantum code is the usual $(2,2)$ four-dimensional toric code on the cellulation.  That is, the corresponding $\ZZ$ chain complex has $Z$-stabilizers on $3$-cells, qubits on $2$-cells, and $X$-stabilizers on $1$-cells.  This $\ZZ$ chain complex obviously admits a lift to the integers, namely the cellular chain complex of the given cellulation using integer coefficients.
Since the cellulation has bounded geometry, the code $B$ is LDPC.

Recall that $H_j(RP^3\times I;\ZZ)=H_j(RP^3)=\ZZ$ for $j=0,1,2,3$.  The nontrivial homology class in $H_1$ can be represented by an $RP^1$ in the $RP^3$ and that in $H_2$ can be represented by an $RP^2$ in the $RP^3$.  Hence, the code $B$ has one logical qubit.

For use later, by Lefshetz duality, $H^2(RP^3\times I;\ZZ)$ is $\ZZ$, and the nontrivial $2$-cocycle is represented by the Poincar\'{e}-Lefschetz dual to $RP^1 \times I$.

We add one additional stabilizer to $B$ to define a quantum code $C$ and corresponding chain complex $\cC$.  
This complex $\cC$ is our example with no sparse lift.
Consider the $RP^3$ at $m=0$.  Choose some representative of nontrivial $H_2(RP^3;\ZZ)$ for this $RP^3$.
The stabilizer we add to define $C$ is the product of all $2$-cells in this representative.
Since we have chosen the $RP^3$ at $m=0$ to have size $O(1)$, the code $C$ is LDPC.
Having added this stabilizer, the code $C$ has no logical qubits.

 We now pick a specific lift $L_Z,L_X$.  We lift all stabilizers of code $B$ to $\ZZZZ$ in the obvious way, using the cellular chain complex of the given cellulation with $\ZZZZ$ coefficients.  We lift the added stabilizer arbitrarily.

Since the chain complex for $B$ admits a lift, the matrix $E$ vanishes on all stabilizers except the added stabilizer.  The image of $E$ on the added stabilizer is a nontrivial $1$-cycle on the $RP^3$ at $m=0$.  This nontrivial $1$-cycle may be taken to be some $RP^1$.

Finding a sparse lift amounts to finding a sparse solution to the equation $E +\partial_Q \delta_Z + \delta_Q \partial_Z=0$.  
Let us refer to this quantity $E +\partial_Q \delta_Z + \delta_Q \partial_Z$ as the ``error".  The error is a map from $Z$-stabilizers to $X$-stabilizers and we say the error ``on a given $Z$-stabilizer" is the image of $E$ acting on a given $Z$-stabilizer.

To get oriented, let us ask how  the error changes if we changes $\delta_Q$ or $\delta_Z$.  The matrix $\delta_Z$ is a map from $Z$-stabilizers to qubits.  If we change some entry in this matrix, i.e., pick a given $Z$-stabilizer and a given qubit, we change the error on that stabilizer by the boundary of the qubit.
This, by changing $\delta_Z$ we can change the error on any $Z$-stabilizer to any other error which is homologous.  Note then that if we set $\delta_Q$ to zero and change $\delta_Z$ we cannot solve \cref{coneq}: the image of the added stabilizer will always be some homologically nontrivial $1$-cycle.

The matrix $\delta_Q$ is a map from qubits to $X$-stabilizers.  If we change this matrix, i.e., pick any given qubit and any given $X$-stabilizer, we add that $X$-stabilizer to all $Z$-stabilizers in the coboundary of the given qubit.

Let us give first a solution of \cref{coneq} before giving an ansatz for all possible solutions.  Pick $\delta_Q$ to be $RP^1 \otimes (RP^1 \times I) $ and $\delta_Z=0$.  The notation here is as follows.  Being a matrix, any $\delta_Q$ can be written as a sum of outer products (the $\otimes$ symbol).  The first factor in the outer product is on $X$-stabilizers, and the second factor is on qubits.  The quantity $RP^1$ is the nontrivial $1$-cycle described above.
That is, we have in mind a specific representative of $RP^1$, namely the image of the added stabilizer, however for shorthand we write this specific representative as $RP^1$.  The second factor can be regarded as a $2$-cochain, and we pick the nontrivial element of $H^2$; we denote this nontrivial element by its Poincar\'{e}-Lefschetz dual $RP^1 \times I$.  Equivalently, this nontrivial $2$-cocycle is in the same class as the product of the nontrivial $2$-cocycle in $RP^3$ with the nontrivial $0$-cocycle in $I$.  
This choice of second factor, $RP^1\times I$, has trivial coboundary, but has nontrivial intersection with the added stabilizer (since $RP^1$ intersects $RP^2$ in $RP^3$).  Note that for the second factor \emph{any} representative in the same class as $RP^1 \otimes I$ will work as a solution, rather than needing a specific representative as we did in the first factor.
So, this is a solution.  Call this solution $\eta_Q$.

Remark: let us discuss why the second factor in $\delta_Q$ should be a cocycle.  Of course, one reason is to solve the equation \cref{coneq}.
However, another way to think about it is that by picking a lift for the added stabilizer, we have also picked a lift for other representatives of $RP^2$: we can obtained other representatives of $RP^2$ (including representatives supported on any given $m$) by taking the given lift of the added stabilizer and adding lifts of other $Z$-stabilizers; that is, the lift of a sum of stabilizers can be taken to be the sum of the lifts, with the terms in the sum of the lifts being taken with arbitrary coefficients $\pm 1$.  So, we seek a solution $RP^1 \otimes w$, for some $w$ which has nontrivial intersection with all representatives of $RP^2$.

Now let us consider all 
possible $\delta_Q$ which could solve \cref{coneq}. Since by changing $\delta_Z$ we can change the error on any $Z$-stabilizer to any other error which is homologous, this question is equivalent to finding all possible $\delta_Q$ such that $E +\delta_Q\partial_Z$ maps every $Z$-stabilizer to something nullhomologous.  We write this as $E+\delta_Q \partial_Z \sim 0$.  Given some solution to the equation $E+\delta_Q \partial_Z \sim 0$, such as 
$\delta_Q=RP^1 \otimes (RP^1 \times I) $, any other solution $\delta'_Q$ can be written as $\delta'_Q=\delta_Q+\kappa_Q$ where
\be
\label{kQe}
\kappa_Q \partial_Z \sim 0.
\ee

We now show that all solutions to \cref{kQe} can be written in the form
\be
\label{allform}
\kappa_Q = \partial_Q M_{QQ} + M_{XX} \partial_Q,
\ee
for some matrices $M_{QQ},M_{XX}$ mapping $2$-chains to $2$-chains and $1$-chains to $1$-chains, respectively.
First, any $\kappa_Q$ of the form \cref{allform} is a solution to \cref{kQe} since $M_{XX} \partial_Q \partial_Z=0$ and $\partial_Q M_{QQ} \sim 0$.
To show that these are all solutions, let $\{v_1,\ldots,v_m\}$, for some $m$, be a basis for $2$-cocycles, using the boundary operator $\partial_Z$ for the code $C$.  Let $w_1,\ldots,w_l$ for some $l$ be some other $2$-cochains such that $\{v_1,\ldots,v_m,w_1,\ldots,w_l\}$ is a basis for all $2$-cochains, i.e., for the columns of $\kappa_Q$.  Since it is a basis, we may decompose
$\kappa_Q=\sum_a x_a \otimes v_a + \sum_b y_b \otimes w_b$ for some $1$-chains $x_a,y_b$.
Note that all $2$-cocycles are trivial, so that $v_a=\partial_Q^T s_a$ for some $1$-cochain $s_a$ where the superscript $T$ denotes transpose: if instead we used the boundary operator $\pZB$ for the code $B$, there is a nontrivial $2$-cocycle but due to the added stabilizer in code $C$ this is not a cocycle for $\partial_Z$.
Hence, $\sum_a x_a \otimes v_a=\sum_a x_a \otimes (\partial_Q^T s_a)$ which is of the form $M_{XX} \partial_Q$.
Since the vectors $\partial_Q^T w_b$ are linearly independent, \cref{kQe} imposes that $y_b \sim 0$ for all $b$, so that $y_b=\partial_Q t_b$ for some $2$-chain $t_b$, and so $\sum_b y_b \otimes w_b=\sum_b (\partial_Q t_b) \otimes w_b$, which is of the form $\partial_Q M_{QQ}$.

\subsection{The Example Has No Sparse Lift}
\label{proof}
From above, we can write $\delta_Q$ in the form
\be
\delta_Q =RP^1 \otimes (RP^1 \times I) + \partial_Q M_{QQ} + M_{XX} \partial_Q.
\ee
Intersect this $\delta_Q$ with the $RP^3$ at some given, large $m$, where we mean intersecting the qubits with the given $RP^3$ so we are taking some subset of the rows of $\delta_Q$.
This intersection then is of the form
\be
\delta_Q(m) =RP^1 \otimes RP^1 + \partial_Q M_{QQ} + M_{XX} \partial_Q(m) ,
\ee
where we have restricted the columns of $M_{QQ}$ to $2$-cells in the given $RP^3$, and restricted the columns of $M_{XX}$ to $1$-cells in the given $RP^3$, and where we define $\partial_Q(m)$ to be a map from $2$-cells in the given $RP^3$ to $1$-cells in the given $RP^3$.
On the right-hand side, the second factor in the term $RP^1 \otimes RP^1$ is $RP^1 \times I$ intersected with the given $RP^3$.
The matrix $M_{QQ}$ contains qubits only in the given $RP^3$.
Through the rest of this subsection, we restrict $M_{QQ},M_{XX}$ in this way.  Also, for brevity, from now on we write $\partial_Q$ rather than $\partial_Q(m)$.

Let us assume, by way of contradiction, that $\delta_Q$ is sparse.  Then, for large enough $m$, the $1$-cells in the image of $\delta_Q(m)$ cannot be close to the $0$ end of the interval $[0,1]$.  Precisely, for any $m'$, for some $m$ which is large, all the $1$-cells in the image of $\delta_Q(m)$ are distance at least $m'$ from the $0$ end of the interval $[0,1]$. This holds because for any given $m'$ there are only $O(1)$ cells within distance $m'$ of the $0$ end of the interval, and so it is not possible to be sparse and to have one of those cells in the image of every large $m$.

In general we may write
\be
\delta_Q(m) =u \otimes v+ \partial_Q M_{QQ} + M_{XX} \partial_Q ,
\ee
where $u$ is any representative in the same class as $RP^1$ and $v$ is any representative in the same class as  the Poincar\'{e}-Lefschetz dual of $RP^1$.
We can change the choice of representative $u,v$ by absorbing it into $M_{QQ}$ and $M_{XX}$.

Let us pick $u,v$ both to both be minimum weight (here weight is Hamming weight) (co)cycles, such that $v$ is supported on the given $RP^3$ and such that
$u$ is supported at least distance $m'$ from the $0$ end of the interval $[0,1]$.
Then, we may also assume that $M_{QQ}$ and $M_{XX}$ vanish if the row is less than distance $m'$ from the $0$ end of the interval $[0,1]$.

Note that $u \otimes v$ defines a matrix which is neither row-sparse nor column-sparse since both $u,v$ have large Hamming weight.  Here a matrix is row-sparse if, for every row, the Hamming weight is $O(1)$, and similarly for column-sparse.

However, it is possible to choose $M_{XX}$ such that $u\otimes v + M_{XX} \partial_{Q}$ is column sparse.  Pick $M_{XX}$ as follows.  For the $i$-th row, if $i$ is \emph{not} in $u$, then that row vanishes.  However, if $i$ is in $u$, then pick that row so that the (co)boundary of that row equals $v+v_i$ for some other $v_i$which is some other representative of $RP^1$.  Pick the $v_i$ so that any given qubit is in at most $O(1)$ different $v_i$.  Then, each row has large Hamming weight, but each column has Hamming weight at most $O(1)$: a given column corresponds to some given qubit, and the nonvanishing entries are the rows $i$ for which that qubit is in $v_i$.
There is some geometry required to show that we can pick the $v_i$ in this way.  However, we omit this since our ultimate goal is not to show that we can pick $M_{XX}$ to make $u\otimes v + M_{XX} \partial_{Q}$ be column sparse, but rather to show that we \emph{cannot} pick $M_{XX}$ and $M_{QQ}$ to make
$\delta_Q(m)$ both row-sparse and column-sparse.

Suppose we choose $M_{XX}$ such that $\partial_X\delta_Q= \partial_X M_{XX} \partial_{Q}$ is not row-sparse, i.e., such that some vertex $v$ has a large number of nonzero entries in the corresponding row $r_v$.  Since $r_v$ is the sum, over edges $e$ in the coboundary of $v$, of the corresponding row $r_e$ of $\delta_Q$, each nonzero entry of $r_v$ must be a nonzero entry of at least one such $r_e$.  Since there are only $O(1)$ edges $e$ in the coboundary of $v$, at least one such $r_e$ must have a large number of nonzero entries, and so $\delta_Q(m)$ cannot be row-sparse, regardless of choice of $M_{QQ}$

So, we may assume that $\partial_X (u\otimes v + M_{XX} \partial_{Q})$ is row-sparse.  The cycle $u$ corresponds to some closed path of edges, $e_0,e_1,\ldots,e_{\ell-1}$, for some $\ell$.  For $\partial_X (u\otimes v + M_{XX} \partial_{Q})$ to be row-sparse, the row of $u\otimes v + M_{XX} \partial_{Q}$ corresponding to edge
 $e_{j}$ must be almost the same as the row corresponding to edge $e_{(j+1) \mod \ell}$, i.e., they may differ in only $O(1)$ entries.  
Since each row has a large number of nonzero entries, this means that there must be some column in $u \otimes v + M_{XX} \partial_{Q}$ such that the column has a non-vanishing entry in some path of edges $e_j,e_{j+1},\ldots,e_{k \mod \ell}$, for some $k$ with $k-j$ large.

We claim then that that column is not sparse in the matrix $\delta_Q(m) =u \otimes v+ \partial_Q M_{QQ} + M_{XX} \partial_Q$.
Adding $\partial_Q M_{QQ}$ to $u \otimes v + M_{XX} \partial_{Q}$ can change that column by adding a boundary.  However, adding that boundary cannot reduce the weight of the column, as if it did then we could shift $u$ by that boundary to reduce its Hamming weight, contradicting the assumption that $u$ has minimum weight; to see this, note that the nonzero entries of the given column are a subset of the nonzero entries of $u$ and so adding that boundary to $u$ would reduce its weight.

\subsection{A Remark on Orientability}
Our construction added a stabilizer acting on a nontrivial $RP^2$ in the boundary $RP^3$.  The reader may wonder whether the crucial property of $RP^2$ was that is was non-orientable or whether it was that $RP^2$ is not the boundary of any $3$-manifold.
We believe that the crucial property is that it is non-orientable.  Suppose we had used a similar construction but instead used a cellulation of $(K^2 \times S^1) \times [0,1]$, where $K^2$ is the Klein bottle.  Suppose the added stabilizer was on $K^2$ which represents a nontrivial homology class in $K^2 \times S^1$.  The Klein bottle is non-orientable but it is the boundary of some $3$-manifold $M$: the Klein bottle is an $S^1$ bundle over $S^1$ and we may take $M$ to be a disc bundle over $S^1$. One might try adding this stabilizer to the stabilizer group by attaching some cellulation of $M$ to the boundary $K^2 \times S^1$ of $(K^2 \times S^1) \times [0,1]$, i.e., considering some code defined by a cellulation of $M$ attached to a cellulation of $(K^2 \times S^1) \times [0,1]$.  This cellulation does indeed add the desired stabilizer to the stabilizer group and it clearly gives a code that has a lift to the integers as it comes from a cellulation.

However, we have added some extra cells by doing this: added $3$-, $2$-, and $1$-cells on the cellulation of $M$ correspond to other added $Z$-stabilizers, qubits, and $X$-stabilizers.  Adding cells is not in itself so serious: the reader might imagine that we are considering some ``stabilization" of the problem: in such a stabilization of the problem, we ask whether we can take a given chain complex over $\ZZ$, i.e., a given quantum code, and then add some extra qubits to that code, with each added qubit $i$ having a single stabilizer (either $Z_i$ or $X_i$) that acts only on that one added qubit, so that there is some sparse lift of that code with the added qubits.  So, one might wonder whether the added cells of $M$ are equivalent, up to local quantum circuit, to a stabilization.  (Further, in some cases, stabilization is equivalent, up to local quantum circuit, to simply choosing some other, coarser cellulation of the manifold.)

However, in this case the added cells do something quite severe, so that adding these cells is likely not equivalent to stabilization up to local quantum circuit.
The added cells kill a certain first homology class $[\alpha]$ of $K^2$ (this class corresponds to the orientation domain wall).  This, for example, changes the number of logical qubits of the code, as the second homology class $[\alpha]\times S^1$ is also killed.
Another effect is that, if we killed this homology class on both boundaries, then
$[\alpha] \times [0,1]$ would give a new nontrivial second homology class, corresponding to some new logical qubit; more precisely, we could modify $[\alpha]\times [0,1]$ by extending it into the added cells on each boundary to give a new nontrivial cycle.

So, while we have not given a proof that this construction with a cellulation of $(K^2 \times S^1) \times [0,1]$ can give a chain complex with no sparse lift, we suspect that it is true, and that a similar proof would work as that used here, so we believe that indeed non-orientability is the crucial property.

\appendix
\section{Every $\ZZ$ Chain Complex Which is Local in One Dimension Admits a Lift (To the Integers) Which is Local in One Dimension}
\label{locallift}

In our example of a chain complex which has no sparse lift, we made the size of the $RP^3$ increase with increasing $m$.
This is necessary, due to a general phenomenon, stated in the title to this subsection.  To explain it in more detail, consider an arbitrary $\ZZ$ chain complex, subject to a certain geometric locality condition as follows.
We have some map from qubits to integers; call these integers ``sites".  We require that each $Z$-stabilizer and each $X$-stabilizer be supported only on a pair of neighboring sites.
In the language of boundary operators, for any column of the boundary operator $\partial_Z$ or any row of the boundary operator operator $\partial_Q$, there is some integer $j$ such that the nonvanishing elements of that column or row correspond to site $j$ or $j+1$.
One may think of this as some one-dimensional lattice quantum system, with possibly more than one qubit per site.

We claim that
any such $\ZZ$ chain complex admits a lift to the integers such that the lifted operators $\tilde \partial_Z$, $\tilde\partial_Q$ obey the same locality property:
for any column of the boundary operator $\tilde \partial_Z$ or any row of the boundary operator operator $\tilde \partial_Q$, there is some integer $j$ such that the nonvanishing elements of that column or row correspond to site $j$ or $j+1$.
Further, the lift has no torsion in its homology or cohomology, and has the same Betti numbers over the integers as the original complex does over $\ZZ$.

To prove this, we first claim that a quantum circuit exists with the following properties.  The circuit consists of two rounds.  The first round is a product of unitaries, each of which is product of CNOT gates, and each of which is supported only on a single site.  The second round is a product of unitaries, each acting on a pair of neighboring sites $j,j+1$, each unitary comprised of CNOT gates, and so that the unitary on sites $j,j+1$ acts on a disjoint set of qubits from that acting on sites $j+1,j+2$.
Finally, 
if we act on the stabilizers with this quantum circuit, the resulting set of $Z$-stabilizers acts on a disjoint set of qubits from the resulting set of $X$-stabilizers, i.e., no qubit is in the support of both some $Z$-stabilizer and some $X$-stabilizer. This circuit ``disentangles" the stabilizers.
Given that this circuit exists, the claim follows.  The image of the stabilizers under the quantum circuit corresponds to a chain complex whose naive lift to the integers is a valid lift (it obeys the requirement that boundary squared is zero), and whose naive lift to integers obeys the claims about torsion and Betti numbers.  Take this lifted complex and then ``undo" the quantum circuit, by acting with a lift of the quantum circuit on the lifted complex.  To describe the lift of the quantum circuit, each CNOT gate applies some row operation on the operator $\partial_Z$, left-multiplying $\partial_Z$ by a matrix which is $1$ on the diagonal and zero everywhere else except it has a $1$ in one off-diagonal entry.  Lift this CNOT gate naively to the integers and left-multiply 
by the products of these lifts to obtain $\tilde \partial_Z$.
Similarly, right-multiply $\partial_Q$ by the inverse of the product of these lifts.

To see that this circuit exists,
consider any site $j$.  The stabilizers (both $X$- and $Z$-stabilizers) supported on sites $j-1,j$ commute with those supported on sites $j,j+1$.  Using standard results on commuting terms which are local in this fashion (for example \cite{bravyi2005commutative}), there is some unitary (indeed, a unitary which is a product of CNOT gates) acting on site $j$ with the property that, after acting with this unitary, the $X$-stabilizers acting on sites $j-1,j$ act on a disjoint set of qubits from the $Z$-stabilizers acting on sites $j,j+1$, and similarly with $X$-stabilizers and $Z$-stabilizer interchanged.  Then, if a qubit on site $j$ is acted on by a stabilizer on sites $j-1,j$ and a stabilizer on sites $j,j+1$, that qubit is either acted on only by $X$-stabilizers or only by $Z$-stabilizers.  A Pauli $X$ or $Z$ operator, respectively, acting on such a qubit is a central element of the interaction algebra of Ref.~\cite{knill2000theory} and so we call them the ``central qubits".  Consider then the set of stabilizers supported on any pair of sites $j-1,j$; other then the central qubits, these stabilizers are supported on a disjoint set of qubits, $S_{j-1,j}$ from those stabilizers supported on sites $j,j+1$ or $j-2,j-1$ so we next apply a quantum circuit acting on qubits in $S_{j-1,j}$ to bring these stabilizer to a form so that $X$- and $Z$-stabilizers act on disjoint sets of qubits from each other.

\bibliography{sparse-ref}

\begin{thebibliography}{3}
\expandafter\ifx\csname natexlab\endcsname\relax\def\natexlab#1{#1}\fi
\expandafter\ifx\csname bibnamefont\endcsname\relax
  \def\bibnamefont#1{#1}\fi
\expandafter\ifx\csname bibfnamefont\endcsname\relax
  \def\bibfnamefont#1{#1}\fi
\expandafter\ifx\csname citenamefont\endcsname\relax
  \def\citenamefont#1{#1}\fi
\expandafter\ifx\csname url\endcsname\relax
  \def\url#1{\texttt{#1}}\fi
\expandafter\ifx\csname urlprefix\endcsname\relax\def\urlprefix{URL }\fi
\providecommand{\bibinfo}[2]{#2}
\providecommand{\eprint}[2][]{\url{#2}}

\bibitem[{\citenamefont{Freedman and Hastings}(2021)}]{freedman2021building}
\bibinfo{author}{\bibfnamefont{M.}~\bibnamefont{Freedman}} \bibnamefont{and}
  \bibinfo{author}{\bibfnamefont{M.}~\bibnamefont{Hastings}},
  \bibinfo{journal}{Geometric and Functional Analysis}
  \textbf{\bibinfo{volume}{31}}, \bibinfo{pages}{855} (\bibinfo{year}{2021}).

\bibitem[{\citenamefont{Bravyi and Vyalyi}(2005)}]{bravyi2005commutative}
\bibinfo{author}{\bibfnamefont{S.}~\bibnamefont{Bravyi}} \bibnamefont{and}
  \bibinfo{author}{\bibfnamefont{M.}~\bibnamefont{Vyalyi}},
  \bibinfo{journal}{Quantum Information and Computation}
  \textbf{\bibinfo{volume}{5}}, \bibinfo{pages}{187} (\bibinfo{year}{2005}).

\bibitem[{\citenamefont{Knill et~al.}(2000)\citenamefont{Knill, Laflamme, and
  Viola}}]{knill2000theory}
\bibinfo{author}{\bibfnamefont{E.}~\bibnamefont{Knill}},
  \bibinfo{author}{\bibfnamefont{R.}~\bibnamefont{Laflamme}}, \bibnamefont{and}
  \bibinfo{author}{\bibfnamefont{L.}~\bibnamefont{Viola}},
  \bibinfo{journal}{Physical Review Letters} \textbf{\bibinfo{volume}{84}},
  \bibinfo{pages}{2525} (\bibinfo{year}{2000}).

\end{thebibliography}

\end{document}